A possible solution to the Dark Matter, Dark Energy, and Pioneer Anomaly problems via a *VSL* approach

Roee Amit
*Division of Biology and Division of Chemistry and Chemical Engineering*
*California Institute of Technology*
*Pasadena, CA 91125*

## Abstract

I apply the equations of motion derived in the accompanying manuscript for the classical approximation of the vsl-path integral to the Newtonian gravitational field in simple geometries. The *vsl* classical-action, a complex quantity in this case, yields modified Euler-Lagrange equations. This, in turn, leads to the emergence of two equations of motions that must be satisfied concomitantly in order to minimize the complex action. The solutions obtained to the doublet equation of motion include the MOND force law, a dark-energy-like omni-present repulsive gravitational force, a pioneer-like anomaly at the solar system level, and additional predictions, which can be verified with either careful observations or via additional probes to the outer solar system. The exercise carried out in this paper exemplifies the explanatory potential of the *vsl*-approach, pointing to a potentially new physics paradigm. Finally, the vsl-approach is not only predictive, but highly falsifiable, an important ingredient of any physics theory.

# I. INTRODUCTION

Newtonian gravitational theory and General Relativity have been two of the most successful theories in the history of physics. The simple $1/r^2$ law, and its General Relativity extension has proven to be a reliable algorithm when used to explain anywhere from local solar-system dynamics to motion on galactic and inter-galactic scales. Yet, as our observational capability has increased with the advent of more advanced telescopes and instrumentation, dynamical anomalies have began to emerge, which may require us to modify these laws. As of today, there are three classes of anomalies that do not conform to the standard canon, and remain empirical facts lacking a suitable explanation. These are the missing mass problem on galactic scales, the dark energy observation on intergalactic or universe scale, and the purported pioneer anomaly at the solar system scale. There are a myriad of theoretical explanations and models proposed, but none that provide a common solution to all three problems.

It is the purpose of the rest of this paper to first broadly review these phenomena, and then show how *vsl*-mechanics provides a uniform formalism by which one can derive all three gravitational anomalies, thus providing a conceptual frame-work with which all of these phenomena can be unified. Indeed, I intend to show that not only does *vsl*-mechanics provide a solution to these anomalies, but a solution for one requires that the others invariably exist. That is, if hypothetically dark energy and the pioneer anomaly were not discovered empirically, then *vsl*-mechanics would unequivocally predict their existence as do or die tests for the theory.

Consequently, I will first derive solutions to the "classical-regime" equations of motion in very specific geometries, which can be solved analytically. As mentioned above these solutions require a thorough extension of Newton's second law, in the sense that the response of a moving particle to a force field is now reflected through the action of two seemingly independent forces: a dominant (large) and a corrective (small) one. In the case of a gravitational field, the dominant *vsl*-force (on almost all scales) component is identical to the force law proposed by Milgrom in his MOND scheme. This force law has been treated in the literature in detail, and it is now well accepted that no matter what the underlying mechanism it describes accurately the flat rotation curves that characterizes spiral galaxies, and the Tully-Fisher relation. The corrective or minor gravitational force law in the *vsl*-mechanics description is the one responsible for a pioneer-type of anomaly at the level of the solar system, and may also serve as a suitable explanation to the mass discrepancy (an anomalous factor of 2) that has plagued the MOND scheme in galaxy cluster models. Finally, despite not having invoked neither general, special relativity, or quantum mechanical constraints, the *vsl*-mechanics solutions results in a repulsive dark-energy type constant force of the order of $\Lambda^{0.5}$ when the source mass approaches zero. The results presented in subsequent pages have the potential to alter in a profound fashion the way we perceive the physical world around us. The *vsl/mqr* concept, whether real or not, has a tremendous explanatory and predictive potential and should be treated seriously as such.

## II. NEWTONIAN GRAVITATION AND OBSERVATIONAL ANOMALIES – FLAT ROTATION CURVES, DARK ENERGY, AND THE PIONEER ANOMALY

The concept of a mass discrepancy or dark matter in observations of celestial bodies was originated by F. Zwicke, when he found that in the coma cluster the virial mass calculated from motion of galaxies in galaxy clusters far exceeded the gravitational mass measured from collected light (Zwicke 1933). However, it was not until the 1970's that the striking observation of flat rotation curves in spiral galaxies, a seemingly sharp disagreement with Newtonian physics, prompted scientists to predict the existence of an unobservable mass termed "dark matter" [1]. That is, mass whose only effects can be observed through gravitational interactions. In addition, the previous observations of Zwicke and contemporary corroborations of those results for other galaxy clusters, showed that the missing mass problem was a problem on several scales, and as a result led to the quick acceptance of the "dark-matter" paradigm. Namely, that the universe is dominated by non-luminous matter and that the nature of this matter may be exotic and outside the scope of the standard model of physics.

Many different ideas have been put forward over the years as to the nature and makeup of the mysterious dark matter. These included anywhere from the mundane (large population of planets, brown-dwarfs, black holes, etc.) to massive neutrinos to exotic versions of heretofore undiscovered supersymmetric particles. The former two have been for the most part ruled out, and at the time of writing of this paper supersymmetric particles have not been discovered as of yet in a laboratory setup. As a result, the state of affairs today from a microscopic perspective is not much different than it was 25 and 30 years ago when the first flat rotation curves were published. That is, aside for astronomical observational data on very large scales (from galactic to the cosmological scales) there are no other experimental evidence for the existence of dark-matter.

In addition, the model used to describe the distribution of the mysterious mass in spiral galaxies is problematic. Since the mass has to be distributed in such a way that the observed flat rotation curves would be produced by Newtonian mechanics, scientists had to assume distributions whose total mass increases linearly with radius. This fact precluded any precise determination of spiral galaxy masses, since rotation curves remained flat to the last measured point and did not show any sign for the Keplerian $1/r^{1/2}$ decline. Moreover, the fitting of the observed data required the assignment of a 3-parameter distribution that implied a fine-tuned rise in the dark-matter concentration with respect to luminous matter in each galaxy where a rotation curve was available. This lack of generality and fine-tunedness was termed the dark-matter conspiracy. That is, the distribution of dark matter is, in a sense, designed in such a way as to precisely generate flat rotation curves in every galaxy where this phenomenon had been observed. Despite these short comings of the dark-matter paradigm, the status of dark-matter as the leading explanation has only been strengthened in recent years due to cosmological (see [2] and references therein) and gravitational lensing (see [3] and references therein) observations that strongly supported this hypothesis.

The reason why the dark matter hypothesis has been steadfast over the last few years despite the lack of progress in determining the nature and content of the missing material is the fact that the alternative for many physicists is unpalatable. If the dark matter hypothesis is wrong, and in fact all the matter that is out there is luminous (with negligible contributions from free-planets, black-holes, brown dwarfs, etc.), it means that Newton's law of gravitation and General relativity are wrong when it comes down to describing motion on galactic and cosmological scales. But, more importantly, not just slightly wrong (the initial predictions of General Relativity for the Precession of Mercury and Solar Eclipse test provided a small correction to the Newtonian prediction) – profoundly wrong. The assumption that Newton's and Einstein's relativity laws could be so wrong seems to be in complete contrast to everything we have learned in the past three centuries of physics at the lab, on earth, and at the solar system level.

Nevertheless, there have been attempts to introduce new physics laws in relation to the dark matter phenomenon. The most notable attempt is Milgrom's MOND. In 1983-4, Milgrom came out with a series of papers ([4-7]) that restated Newton's second law to account for a special acceleration scale he termed $a_0$ as follows:

$$f = ma\mu\left(\frac{a}{a_0}\right) \tag{1}$$

where $\mu$ is some function whose limits are given according to:

$$\mu\left(\frac{a}{a_0}\right) = \begin{cases} 1 & \text{for } a \gg a_0 \\ a/a_0 & \text{for } a \ll a_0 \end{cases} \tag{2}$$

therefore, in the limit of low accelerations Newton's second law should be quadratic and approach the following form in the presence of a gravitational potential:

$$\frac{a^2}{a_0} = \frac{\partial \phi}{\partial r} \tag{3}$$

where $a$ is the acceleration, $a_0$ is some constant, and $\phi$ is the gravitational potential. Assuming that the Newtonian version of the centripetal law still holds, we then have for a point mass attractive potential:

$$v^4 = GMa_0 \tag{4}$$

which describes asymptotically a flat rotation curve. In addition, he specified candidate mathematical functions for $\mu$, which provides a transition from Newtonian to Mondian dynamics, and as a result the full MOND force law for a gravitational field takes the form (for one particular function):

$$\frac{a^2/a_0}{\sqrt{1+\left(a/a_0\right)^2}} = \frac{\partial \phi}{\partial r} \tag{5}$$

This formula, and the MOND paradigm, much to the chagrin of the physics community had been remarkably successful. In fact, it clearly has withstood the test of time as a viable alternative to the dark-matter hypothesis. Over the years, predictions that resulted from the MOND hypothesis particularly for Low Surface Brightness galaxies and dwarf spiral galaxies have panned out. As of today, MOND (see MOND pages website: http://www.astro.umd.edu/~ssm/mond/litsub.html, and [8]) was used to fit successfully previously published non-perturbed spiral galaxy rotation curves, which amount to nearly 100 such measurements. Fitting the rotation curves, in it by itself, is not a great theoretical achievement, particularly if one develops a model that necessitates three fitting parameters as in the dark-matter case (two parameters for dark matter and one for luminous mass), but MOND's success is rooted in the fact that it necessitated only one fitting parameter: the luminous mass to light ratio ($a_0$ is a prediction as pointed in [9]), indicating that the mass that drives the motion is distributed in the same way as the light emitted by the galaxy, and therefore there is no need for "dark matter" to explain the motion.

Moreover, MOND's phenomenological successes have extended much further than predictions of spiral galaxies rotations curves, and have been able to reproduce other astrophysical observations attributed to the presence of dark matter with just the luminous mass. Particularly, we note MOND's relative success in explaining the virial mass discrepancies of small groups of galaxies, rich clusters, globular clusters and molecular clouds, and possibly super-clusters as well. [8]. One notable failure, which may or may not indicate the demise of MOND is its lack of success in determining correctly the masses of large cluster cores due to estimates of radial temperature profiles which disagree with current observations [10]. That being said, MOND's wide array of successes has prompted the famous late astrophysicist J.N. Bachall to remark: "Something deep is right about MOND, if only to describe in a succinct way a number of a priori surprising regularities in the data for galactic systems" [11].

However, there is a problem. MOND, as stated by Milgrom himself clearly, is not a theory. It is an hypothesis. That fact has lead to the viewing of this successful approach by the community as nothing more than a fortuitous curiosity, rather than a flashing red-light indicating that there may be something very wrong with physics as we know it. As a result, MOND does not make predictions on the cosmological or microscopic scales, and is limited to making phenomenological observations on galaxies and galaxy cluster scales. Due to this fact and the gravitational lensing data supporting the existence of dark-matter, Milgrom, Bekenstein and their collaborators have focused their efforts on finding the theoretical source for MOND from a modification to Einstein's General Relativity, which reduces to MOND in the appropriate limit. Only recently, a consistent covariant theory has finally been developed [12]. However, it still remains to be seen whether or not this theory will be accepted as a consensus source for MOND, and either way it does not provide for a microscopic explanation for the MOND phenomenology. The lack of a coherent theoretical source for MOND over the years has left this idea on the fringes of modern physics thought, with only a handful of researchers actively working on it or similar approaches.

In recent years, two more classes of gravitationally anomalous behavior have been discovered. The first, is what is termed today dark energy, while the second (which at present is yet to be confirmed) is called the pioneer anomaly. Much have been written and said about dark energy (see for example [13]) which is considered to be the dominant form of gravitational mass density in the present day universe. Dark Energy corresponds to a repulsive gravitational force that is omni-present (i.e. non local), constant, and time-independent. On the macroscopic scale it is often associated with Einstein's cosmological constant, but the conventional wisdom usually treats its source as due to fluctuations in the zero-point energies of particle and fields or vacuum energy. The dark energy paradigm emerged from a series of complicated data analysis based on varied astronomical data and direct observational evidence that suggest that the rate of scale factor expansion of the universe is not decreasing but in fact increasing (i.e. the universe's expansion is accelerating!). This indicates that there is a "repulsive" gravitational force at work, and that the best fit model to the total mass density in the universe is one with a non-zero cosmological constant density factor of $\Omega_{\Lambda 0} \sim 0.7$ (see [2, 3] and references therein). Whatever the mechanism for dark-energy, there is a consensus in the physics and astrophysics community that it will emerge from a new-type of physical principle.

Both the dark matter and dark energy paradigms are gravitational phenomenon that occur on very large scales, and are therefore hardly accessible to studies at the laboratory. Solar system level studies of Newtonian gravitation have been carried out to an ever increasing accuracy by the man-made probes that have been launched to explore the outer reaches of the solar system over the past 35 years. Two of the earlier missions were carried out by the pioneer 10 and 11 probes, whose purpose was to survey Jupiter (pioneer 10) and Saturn (pioneer 11). At the conclusion of their survey the spacecrafts have been continuing on a direct path out of the solar system towards interstellar space. Presently, they are located at a distance of approximately 70 AU from the sun (a little less than twice the distance of Pluto).

It has become apparent in recent years that despite what was originally thought, the pioneer mission is not really over. In a landmark paper [14] reported an anomalous force in the direction of the sun that is acting on the pioneer spaceships. In other words, these crafts are closer to the sun than they should be according to Newtonian physics. Anderson *et al.* judiciously explored all possible systematic causes for this anomaly and concluded that none were found. In their discussion, they supported the notion that a systematic error may be behind the anomaly (either gas leakage or some heat loss mechanism), but also pointed out elegantly that neither of the missing-mass theories (i.e. MOND, dark matter paradigm, etc.) can account for the anomaly in a consistent/coherent fashion. The anomalous force signature occurs for both spacecrafts (despite their independent missions) and is measured to be a constant – independent of the distance to the sun. If this observation turns out to be robust by continued analysis of the mission's data, and by the occurrence of similar anomalies in other probes, it will require as in the previous two cases the invocation of a new physics principle to account for the anomalous acceleration.

In this section, three gravitational phenomena were reviewed that occur at three different scales (solar system, galactic, and universe) and together or separately may require a new physics paradigm. Moreover, all three anomalies seem to emerge as a result of unrelated physics based on current leading paradigms. Namely, the missing mass or dark matter phenomenon has no baring on the dark energy observation, and neither MOND nor the dark matter paradigm can explain the pioneer anomaly (if confirmed to be true). That being said, the constant that characterizes the transition to small accelerations in Milgrom's MOND scheme appears in all phenomenon and may be a common thread that is used to deduce the underlying physics. Namely, the repulsive acceleration induced on a particle due to the cosmological constant component is:

$$a_0 \cong c\Lambda^{1/2} \tag{6}$$

and the value of the pioneer probes' anomalous acceleration is measured to be $(8.74\pm1.33)\times10^{-8}$ cm/sec$^2$ which is about $6a_0$. In other words, MOND may not only be a model that described galactic rotation curves, but also hint at the underlying physics which generates that other gravitational anomalies as well.

### III. THE *VSL*-MECHANICS APPROACH TO NEWTONIAN GRAVITATION

In deriving the classical equations of motion, we must first generate the new *vsl* "classical" action, and from that utilize the traditional calculus of variation techniques in order to derive the equations of motion. We, therefore, insert the Newtonian gravitational term into the classical Lagrangian density Γ (equation 62 – in the accompanying manuscript), and obtain:

$$\Gamma^{vsl} = \left(1 + i\frac{E_u}{16\pi\hbar G}|\nabla\phi|\right)\frac{|\nabla\phi|^2}{8\pi G} \tag{7}$$

This equation can be rewritten in complex polar coordinates as follows:

$$\Gamma^{vsl} = \frac{|\nabla\phi|^2}{8\pi G}\sqrt{1 + \left|\frac{E_u\nabla\phi}{16\pi\hbar G}\right|^2}\exp\left(i\tan^{-1}\left(\frac{E_u\nabla\phi}{16\pi\hbar G}\right)\right) \tag{8}$$

Using the following definition:

$$\frac{1}{a_0} = \frac{E_u}{16\pi\hbar G} \tag{9}$$

We rewrite equation (65) as follows:

$$\Gamma^{vsl} = \frac{|\nabla\phi|^2}{8\pi G}\sqrt{1 + \left|\frac{\nabla\phi}{a_0}\right|^2}\exp\left(i\tan^{-1}\left(\frac{\nabla\phi}{a_0}\right)\right) \tag{10}$$

In order to evaluate the physical meaning of the above complex Lagrangians, we assume that as usual motion will be defined by the path that minimizes this action. The difference is that this action is complex. From a mathematical stand point the operation is identical, however, from a physical stand point a complex action is a non-intuitive concept, particularly for the classical regime.

Therefore for the Newtonian gravitational field, taking the variation of equation (10) (and putting this into matrix form) we get:

$$\nabla \bullet \begin{pmatrix} F_\Lambda \\ F_{mond} \end{pmatrix} = 4\pi G \begin{pmatrix} \rho_R \\ \rho_I \end{pmatrix} \begin{pmatrix} \cos(\alpha) & \sin(\alpha) \\ -\sin(\alpha) & \cos(\alpha) \end{pmatrix} + \frac{\partial \alpha}{\partial r} \begin{pmatrix} F_{mond} \\ -F_\Lambda \end{pmatrix} \tag{11}$$

where,

$$F_\Lambda = \nabla\phi \frac{1 + \frac{3}{2}\left|\frac{\nabla\phi}{a_0}\right|^2}{\sqrt{1 + \left|\frac{\nabla\phi}{a_0}\right|^2}} \tag{12}$$

$$F_{mond} = \frac{\frac{|\nabla\phi|^2}{2a_0}}{\sqrt{1 + \left|\frac{\nabla\phi}{a_0}\right|^2}} \tag{13}$$

$$\alpha = \tan^{-1}\left(\frac{\nabla\phi}{a_0}\right) \tag{14}$$

and $\rho_R$ and $\rho_I$ are the corresponding real and imaginary mass terms which are yet to be defined. The striking feature of equation (13) is that it is identical to the left hand side of Milgrom's MOND force law (5) (up to a factor of 2) as defined in his Lagrangian version of his scheme called AQUAL [7]. While the MOND force law was implemented as a singular force law used to modify Newton's second law or gravitational force for extremely low accelerations or gravitational forces, in the *VSL* scenario it emerges naturally as one of the two forces in the doublet force law. In this case, the MOND force law corresponds to the dominant part of the doublet of eigenforces, while the other or corrective force (referred to as $F_\Lambda$) will be shown subsequently to be responsible together with the dominant $F_{mond}$ for the dark energy and pioneer anomaly solutions.

Interestingly, Milgrom [15] has shown that the MOND force law cannot be derived from a real or classical-type action. This is mainly due (as can be easily checked by making $E_u$ complex in equation (27) of the accompanying manuscript) to a mathematical constraint that, for real Lagrangians containing a MONDian $|\nabla\phi|^3$ term, generates equations of motions that predict special physics at high accelerations, contrary to empirical evidence. The *vsl* action is complex, and the emergence of MOND is therefore necessarily accompanied by an adjoint force equation. Thus, the complex *vsl* doublet equation of motion is the first such theory to derive the successful MOND scheme from first principles.

The *vsl*-approach treats only the force part of the lagrangian, and does provide a receipt as to how to handle the inertia or mass term. Essentially, what we have above is just the left hand side of the equations of motion, whereas the right-hand side corresponding to the inertia or mass term is yet to be defined. The simplest case, which can be handled precisely, is the one where the mass term is 0 – will be handled in the next section, while the more-complex mass-term derivation will follow in the subsequent section with precise analytical evaluation at the low and high *vsl* asymptotic regimes.

## IV. *T*HE DERIVATION OF A DARK ENERGY-LIKE FORCE

One of the simplest cases to handle, and solve for the doublet equation of motion precisely is the one where the source mass approaches zero. In this case by examining $F_\Lambda$ or by setting the right hand side of equation (12) to 0, we notice that there are two possible solutions (see also the general discussion and equations (68) and (69) in the accompanying manuscript):

$$\nabla \phi = 0 \tag{15}$$

$$\nabla \phi = -i \frac{2}{3} a_0 \tag{16}$$

where the top solution is obviously the trivial solution, which we would expect in such a case. The bottom solution, however, leads to the following stark conclusion: that the dominant or what was termed the MOND force (13) is now non-zero, repulsive, and takes on the value:

$$F_{mond} = -\sqrt{\frac{1}{3}} a_0 \tag{17}$$

This force is a non-local, omni-present, and independent of scale and position. It is repulsive, in the sense that a single particle in an empty universe would feel this acceleration acting on it, pulling it away from an observer. Indeed, when compared to equation (6), we can reformulate this force in terms of the cosmological constant, and as a result the gravitational energy that is associated with this term is indeed just $\Lambda$.

The *vsl*-mechanics doublet force equation generates a non-trivial force solution that corresponds to a "non-zero" ground-state of accelerated motion. All massive particles respond to this constant force, which according to *vsl*-mechanics is nothing more than a consequence of the multiple quantum reality condition. Thus, dark energy is indeed a quantum phenomenon, yet not one which is associated with the vacuum energy. Instead, it is related to the fact that in a *vsl*/*mqr* universe with a minimal length scale and where the planck parameter varies according to some distribution, such a force must inevitably exist as to reflect these constraints. It is important to stress that this non-local force is an inevitable consequence of the quantum *vsl* assumptions, and therefore corresponds to a powerful limitation on the theory. That being said, since a force such as this has been discovered in our universe, this stringent condition renders the *vsl*-explanation a suitable physics paradigm as to the underlying nature of the dark energy force.

Empirically the cosmological constant or dark energy force is a source for apparent paradoxes. If one were to invoke the standard model of physics, counting over the zero-point energies (the currently presumed source for the dark energy force) would result in an unphysical and astronomical value for $\Omega_{\Lambda 0}$. The flip-side of this assessment is that in order to have a cosmological constant density parameter term of 0.7, the value of $\Lambda$ must be of O($10^{-120}$) a ridiculously low and fine tuned number. Moreover, unless we assume that $\Lambda$ evolves with time (which is inconsistent with the current formulation of general relativity), invoking at present a significant $\Omega_{\Lambda 0}$ term implies that we live in a very special time where $\Omega_{m0} \sim \Omega_{\Lambda 0}$. That is, when the universe was one tenth its present size $\Omega_{\Lambda 0} \sim 0.003$, and in the future we should expect an exponential expansion of the universe akin to some of the inflationary scenarios described for primordial time. Presently, according to standard dogma, we are observing a universe that has just recently began to expand with a significant cosmological constant term, and thus live in a very special epoch. All this makes the underlying physics that generates dark energy a genuine mystery, and one which challenges are most cherished paradigms and dogmas.

The *vsl*-mechanics derivation and underlying meaning for the repulsive dark energy force naturally alleviates these paradoxes, eliminating the need to assume a special fine-tuned value or period in history, and thus provides a dark-energy paradigm that does not conflict with the cosmological principle. The fine-tuning problem is alleviated by the fact that the *vsl*-mechanics or MOND $a_0$ parameter is a numerical constant of order $10^{-10}$ m/sec$^2$ that translates (because of the way it is numerically defined) to a very small $\Lambda$. There is nothing, special or fine-tuned about the value of $a_0$, and it is altogether not assumed within the context of the present theory to be related to vacuum energies. The "special-time" paradox, as will be shown below, is alleviated by the fact that *vsl*-mechanics inherently requires $a_0$ to vary over the life-time of the universe, rendering the dark energy force evolvable. As a result, there is nothing special about the present day value of $\Lambda$, and more importantly its relative size as compared with the mass and radiation density parameters. This is at present inconsistent with general relativity (GR), yet it is obvious that the *vsl*-approach would require a concomitant modification to GR, and therefore does not represent a problem at the present time.

In order to consider the consequences of temporal-evolvability of $a_0$, we invoke equations (9) and (17) to get:

$$F_{DE} \propto -a_0 = -\frac{16\pi g\hbar}{E_u} = -16\pi \left(\frac{G\hbar}{c^3}\right)\left(\frac{c^2}{V_u}\right) \tag{18}$$

where $V_u$ is the universon volume as defined by equation (63) of the accompanying manuscript. This can be rewritten as follows:

$$F_{DE} = -16\pi \frac{(cl_p)^2}{V_u} \tag{19}$$

where $l_p$ is the planck length, considered in *vsl* or DSR [16-18] context as the "absolute zero of length", and is a constant that is measured at the same value by all observers no matter the scale. Moreover, since it has been observed that:

$$a_0 \cong cH_0 \tag{20}$$

a fact which is consistent with the *vsl* approach, and using simple dimensional analysis, we can assume that the present day universon volume is therefore given by:

$$V_u \cong l_p^2 \frac{c}{H_0} \tag{21}$$

which is nothing more than the universon version of the expansion of the universe concept. This leads to a natural, and consistent cosmological model that is based on the *vsl*-principle. That is, the value of the cosmological constant decreases over time in a manner proportional to the expansion parameter of the universe. Therefore, what we have is a scenario where all the density parameters are time evolvable, and their ratio at a given point time depends on their particular time dependence. The universe in this scenario has always had an accelerating component, even though in the distant past there may have been a period where it was not the dominant component. In the context of this work, lacking a *vsl*-cosmological model, what we can say about the future of the universe is that it is the jerk that is decreasing and will approach zero as the universe continues to expand.

Equation (21) has further ramifications for the over-all evolution of the average-value of the speed of light and planck parameters. Since *vsl*-mechanics implicitly assumes that both the planck length ($l_p$) and the universon energy ($E_u$) parameters are discrete, time, and scale independent constants (consistent with the DSR approach), one therefore must deduce from equation (21) that:

$$c \propto \sqrt{H_0} \tag{22}$$

$$G\hbar \propto (H_0)^{\frac{3}{2}} \tag{23}$$

Thus leading to the following conclusion: the *vsl*-assumption combined with the observed expansion of the universe since the big bang implies that in order to conserve the planck length and universon energy constants, it is the universon volume which must expand. In other words, the quantum or planckian expression of the expansion of the universe manifests itself in the time-dependent increase of each individual universon's characteristic volume – from a planck volume to present day size. Note, that this expansion of "volume" is highly reminiscent of an adiabatic expansion of an ideal gas. Moreover, the increase in this volume necessarily requires the concomitant decrease in the value of the speed of light parameter, which functions in this case as a "universon temperature". Thus, the expansion of the universe can now be seen as not only as a result of adiabatic processes that take place on matter, but also on the underlying background which must also expand adiabatically with its own thermodynamic sense. This implies that the early universe was very hot (not only in the conventional sense), which led to an inflationary-like scenario characterized a much larger speed of light parameter as compared with the present day, which in turn allowed for the homogenization of the background observed in the CMB today. This was indeed the underlying motivation for the original *vsl* idea suggested by [19] that was designed as an alternative to the standard inflationary scenario. Moreover, a time dependent speed of light and planck constant parameters suggest that the dimensionless fine structure constant has evolved over time as well, a suggestion which is supported by some preliminary evidence [20, 21], and may play an important future role as an empirical test for the *vsl*-idea. Finally, the time variation of the planck constant and the gravitational constant can have additional ramifications for structure formation (galaxies, clusters, voids, etc.) and other primordial phenomenon that at present are difficult to simulate.

It is important to note that the cosmological ramification derived above require a *vsl*-type of cosmological model, which in turn will rely on a *vsl*-version of GR in order to generate reliable predictions. All these are outside the scope of the present work. Moreover, further conclusions from the above equations are sketchy at best suggesting that underlying *vsl*-mechanics there is universon-field theory and universon thermodynamics that puts precise values and predictions on these observations.

## V. THE MASS TERM

In order to derive the right hand side (mass term) of the *vsl*-doublet equation of motion (equation (80)), one approach is to invoke a *vsl* extension of the equivalence principle, and from it derive the proper form of the mass term. The other is to use empirical knowledge to deduce (at least) the asymptotic behavior of this term. In the following, we will derive the mass term as a combination of both strategies, while keeping in mind that neither method will provide a sound theoretical approach for this result. This derivation will complete the *vsl*-mechanics description for non-relativistic gravitational fields and in the process derive the behavior that is associated with solar systems, a pioneer anomaly, and galaxies – dark matter.

In order to do that, and after having shown that Milgrom's MOND field equation emerges as one of the *vsl* doublet eigenforces, it will be preferable to return to a non-polar presentation of the equations of motion in order to derive the rest of the results. Thus, the simplest presentation of the *vsl* doublet is as follows (using equation (7) for the action):

$$\nabla \cdot \begin{pmatrix} \nabla \phi \\ \frac{3}{2} \frac{|\nabla \phi^2|}{a_0} \end{pmatrix} = 4\pi G \begin{pmatrix} \rho_R \\ \rho_I \end{pmatrix} = 4\pi G \begin{pmatrix} f(\rho) \\ g(\rho) \end{pmatrix} \qquad (24)$$

Where the top force corresponds to the real force (or $F_A$ in the polar presentation), while the bottom force is the imaginary component or ($F_{mond}$ in the polar presentation).

There are many choices for what the mass term could take. The simplest would be to assume the following:

$$\nabla \bullet \left( \frac{3}{2} \frac{|\nabla \phi|^2}{a_0} \nabla \phi \right) = 4\pi G \rho \binom{1}{1} \quad (25)$$

that is, a Newton-like inertia response for both forces. In order to evaluate the relative importance of each term we must first define the two asymptotic regimes: The weak and strong *vsl* limits. In the former, we assume that the *vsl*-effect is nothing more than a perturbation on the standard Newtonian lagrangian. Hence, we define it as follows:

$$\frac{E_u \nabla \phi}{16 \pi \hbar G} = \frac{\nabla \phi}{a_0} << 1 \quad (26)$$

note, that this limit corresponds to the extreme MOND regime in Milgrom's scheme, whereas in *vsl*-mechanics this corresponds to the first correction to Newtonian mechanics. The strong *vsl*-limit is likewise defined as:

$$\frac{E_u \nabla \phi}{16 \pi \hbar G} = \frac{\nabla \phi}{a_0} >> 1 \quad (27)$$

which appropriately corresponds to the classical or the Newtonian limit in the MOND scheme.

Typically when one performs a perturbation in physics, the result of the operation induces a small correction to the standard law. In the case of the weak *vsl*-limit, which at the Lagrangian level is defined as a perturbation, this does not occur. In fact for this limit the *1/r* force is dominant. By dominant, we mean that the velocity that a test particle attains due to this field is mostly due to the MONDian force of the doublet. The reason for this is as follows: in the weak *vsl*-regime the smallness of the *vsl* term implies that $a_0$ would be much larger than the field gradient. This therefore means that any inertia response will be dominated by this force. An intuitive way to understand this result is to recognize that for circular orbits in particular, this term implies that one can generate large orbital velocities when condition (26) holds, while the actual "gravitational force-field" is small. The paradox is that according to this scheme an infinitesimal $E_u$ at the quantum scale essentially dominates motion on all classical scales. This is a counterintuitive result, which demands further scrutiny. Indeed, if we slowly crank down the universon energy to 0, the inertia effect of the MOND force becomes further dominant. For example, if $a_0$ were a number on the order of 1 m/s$^2$ planetary motion in the solar system would behave like a 1/r law, and result in a flat rotation curve for the planets (and probably no planets, and no sun to begin with). From the point of view of

*vsl*-mechanics the situation where $E_u$ is identically 0 is an extremely unstable state with regard to gravitational motion. This is, again a "fine-tuning" situation, which is unpleasing from a theoretically esthetic perspective. A non-zero universon energy provides for a stabilizing mechanism, which generates a hierarchy of gravitational force effects on motion on all scales considered non-quantum in the *vsl*-approach. This is indeed what is observed in our universe, as this hierarchy implies a weak *vsl*-effect for small forces (and large distances from the gravitational source), and an increasing *vsl*-effect for smaller distances and large forces. In fact, the strong *vsl*-regime means in the case where condition (27) holds, the typical $1/r^2$ force dominates the motion. This is precisely what is observed from clusters to galaxies in the weak limit-*vsl* limit to solar systems at the strong *vsl* limit.

So how is it possible to differentiate between the classical Newtonian scenario, the MOND scenario, and what is derived from the *vsl*-approach? The answer to this is simple, in Newtonian physics and MOND scenario there is a single force, while in the *vsl*-description the dominant force is accompanied by a minor corrective force. In the weak-*vsl* limit the $1/r^2$ force is present and perturbative in its effects on motion, while in the strong *vsl*-limit the reverse is true. From an empirical perspective, we do not as of yet have the capability to measure perturbative effects of additional forces on extra-galactic rotation curves. However, in the strong *vsl*-limit (our neighborhood) we do indeed. In the context of equations (25) and (26), *vsl*-mechanics predicts a perturbative $1/r$ force at the solar system level. This prediction (see green dash-dot line in figure 1 and discussion in [22]) is a $10^{-5}$ effect at Mars' orbit, which is 4-5 orders of magnitude larger than our current very precise knowledge of the latter's orbit [22]. This means that a $1/r$ force cannot be the aforementioned perturbative effect, and the choice for the mass term in equation (25) cannot be correct for all scales. That being said, the pioneer anomaly suggests that such a perturbative force does exist, and the challenge therefore becomes to construct a mass term, which describes the correct weak *vsl* regime, along side a Newtonian force accompanied by a perturbative constant pioneer anomaly force for the strong *vsl*-regime.

The *vsl*-approach utilized for the static gravitational field is not necessarily applicable to the mass term. As can be checked easily, if we were to use the same operation on the "the full Newtonian Gravitational Lagrangian":

$$L = \frac{|\nabla \phi|^2}{8\pi G} + \rho \phi \tag{28}$$

we would derive equations of motion that result in $1/r^2$ forces for both $F_{mond}$ and $F_A$. This is obviously false, but also does not fit well with the *vsl*-approach. Indeed, what we need here is the appropriate inertia response to the force doublet, and a blind implementation of the *vsl*-operation does not fit the bill in this case.

It is not trivial to generate a natural extension to Newton's law when deriving the mass term. The first assumption is that for small $E_u$, the mass term for both forces in the doublet indeed takes on the Newtonian value as can be expected from a weak-*vsl* regime. For the weak *vsl* limit, we assume that the force doublet replaces the Newtonian F in the second law, while the mass or inertia response remains the same, or in this case being applied as unitary doublet in the same sense to both forces as shown in equation (25). This choice for the mass term is sensible from a physical perspective, and also fits the empirical data well as discussed above.

In the strong *vsl*-limit, we have to make assumptions on the behavior of the mass that we do not need to make in the weak-*vsl* limit. In this case, lacking a good physical reasoning for a given choice we require from the strong *vsl*-limit to produce a $1/r^2$ force as the dominant contribution to the doublet, and a secondary pioneer-anomaly constant force for the other force. One way to construct this coherently, while having a continuous evolution of both the force and mass doublets from the weak-*vsl* limit is to assume the following for the *vsl*-mechanics equation of motion:

$$\nabla \bullet \begin{pmatrix} \nabla\phi \\ \dfrac{3}{2}\dfrac{|\nabla\phi^2|}{a_0} \end{pmatrix} = 4\pi G\rho \begin{pmatrix} \cos\alpha \\ \sec\alpha \end{pmatrix} \tag{29}$$

where $\alpha$ is defined in equation (14). This equation can be expressed in complex polar coordinates, thereby allowing us to view the full *vsl*-doublet equation of motion in terms that include the MOND formalism as follows:

$$\nabla \bullet \begin{pmatrix} F_\Lambda \\ F_{mond} \end{pmatrix} = 4\pi G\rho \begin{pmatrix} \cos(\alpha) \\ \sec(\alpha) \end{pmatrix} \begin{pmatrix} \cos(\alpha) & \sin(\alpha) \\ -\sin(\alpha) & \cos(\alpha) \end{pmatrix} + \frac{\partial\alpha}{\partial r}\begin{pmatrix} F_{mond} \\ -F_\Lambda \end{pmatrix} \tag{30}$$

In order to check that equation (30) satisfies the requirements for gravitational phenomena from the weak to the strong *vsl*-limit, we graph both the doublet forces as a function of normalized distance in figure 1 (where $R_m = \sqrt{GM/a_0}$). Here, we clearly view the behavior of both forces from the strong to the weak *vsl* limits. In the strong *vsl*-limit $F_{mond}$ (blue dashed line) behaves like a Newtonian $1/r^2$ force, while $F_\Lambda$ (red solid line) is approximately a constant of the order of $a_0$. As the ratio approaches one the forces become nearly equal in terms of contribution to motion, and then further evolve to the familiar $1/r$ force for $F_{mond}$ and $1/r^2$ force for $F_\Lambda$ in the weak-*vsl* limit.

In order to drive the pioneer anomaly force, we must derive analytically the first perturbations to the *vsl* force law the strong *vsl* limit. Therefore, when condition (27) holds we have:

$$\nabla \bullet \begin{pmatrix} \nabla\phi_\Lambda \\ \dfrac{3}{2}\dfrac{|\nabla\phi_{mond}|^2}{a_0} \end{pmatrix} = 4\pi G\rho \begin{pmatrix} \dfrac{a_0}{\nabla\phi}\left[1 - \dfrac{1}{2}\left|\dfrac{a_0}{\nabla\phi}\right|^2 + \ldots\right] \\ \dfrac{\nabla\phi}{a_0}\left[1 + \dfrac{1}{2}\left|\dfrac{a_0}{\nabla\phi}\right|^2 + \ldots\right] \end{pmatrix} \tag{31}$$

For the strong *vsl* limit (as figure 1 shows), we can assume the following:

$$\nabla \phi \equiv \nabla \phi_{mond} + \nabla \phi_\Lambda \cong \nabla \phi_{mond} \qquad (32)$$

Which is nothing more than the Newtonian gravitation law in this regime. Furthermore, assuming a circularly symmetric orbit, or 1-D linear system, and setting:

$$\rho' = \rho/3 \qquad (33)$$

then the following holds:

$$\nabla \cdot \begin{pmatrix} \nabla \phi_\Lambda \\ \nabla \phi_{mond} \end{pmatrix} \cong \nabla^2 \phi_N \begin{pmatrix} \frac{3a_0}{\nabla \phi_n} \\ 1 \end{pmatrix} \qquad (34)$$

where $\nabla \phi_n$ corresponds to Newtonian gravitation field satisfying the poisson equation. This therefore implies the following for a 1-D system:

$$\begin{pmatrix} \nabla \phi_\Lambda \\ \nabla \phi_{mond} \end{pmatrix} \cong \begin{pmatrix} 3a_0 \ln \nabla \phi_n \\ \nabla \phi_n \end{pmatrix} \qquad (35)$$

which describes the gravitational force at the solar system level (strong *vsl* limit), accompanied by a corrective force effect that is of the same order as the observed pioneer anomaly (up to a factor of two pi). Note, the result obtained in equation (35) does not apply to closed orbits, and can be used as an approximation only for a particle moving on a linear (or hyperbolic) escape trajectory such as the pioneer probes. For closed orbits, one needs to solve the appropriate Poisson equation to get the solution. However, by inspection one can tell that the anomalous force for circular orbits will be a constant force on average (not logarithmic as for the escape trajectory) of the order of $a_0$. Interestingly, [22] listed in his analysis the degree of accuracy to which we know the inner planets orbits, a number which hovers around $a_0$, in agreement with the result obtained here.

Using the same approach we can derive the first order perturbation to the *vsl* force law at the solar system level, thus generating a prediction as to where will a deviation from the constant pioneer anomaly force will appear. Returning to the strong *vsl* limit, equation (31) becomes:

$$\nabla \cdot \begin{pmatrix} \nabla \phi_\Lambda \\ \nabla \phi_{mond} \end{pmatrix} \cong \nabla^2 \phi_N \begin{pmatrix} \dfrac{3a_0}{\nabla \phi_n} - \dfrac{3}{2}\left[\dfrac{a_0}{\nabla \phi}\right]^3 \\ 1 + \dfrac{1}{2}\left[\dfrac{a_0}{\nabla \phi}\right]^2 \end{pmatrix} \tag{36}$$

and in the particular special geometry of linear 1-D motion, we have:

$$\begin{pmatrix} \nabla \phi_\Lambda \\ \nabla \phi_{mond} \end{pmatrix} \cong \begin{pmatrix} 3a_0 \ln \nabla \phi_n + \dfrac{3}{4} a_0 \left|\dfrac{a_0}{\nabla \phi_n}\right|^2 \\ \nabla \phi_n - \dfrac{1}{2} a_0 \left|\dfrac{a_0}{\nabla \phi}\right| \end{pmatrix} \tag{37}$$

for the first order correction terms. Note, that the first order correction term to the dominant force in the doublet is smaller than the $0^{th}$ order term in the minor force. This implies that the first correction to the total solar system gravitational force, will manifest itself as an anomalous $r^2$ force which will be interpreted as an r-dependence for the pioneer anomaly. At the present time the current positions of the pioneer probes (R~100 AU) means that this correction has a value of about $10^{-4}$ $a_0$ – or about 4 orders of magnitude smaller than the measured value of the pioneer anomaly. At about a 1000 AU, where this number will reduce to $10^{-2}$ – it is likely that we could be able to measure this correction directly. Note, that this correction emerges from the MOND scheme as well, and it is this term that has precluded MOND from claiming success for the pioneer anomaly. In the *vsl* derivation this term is placed in its proper context as the first corrective term to appear aside for the standard Newtonian and constant pioneer anomaly forces. It is important to stress that this correction is a prediction. Again, this can serve as a definitive test for the *vsl*-approach, as higher order corrections to the Newtonian prediction (starting with a $r^2$ force) of the motion of the pioneer probes will become increasingly dominant as the gravitational field acceleration will continue to decrease towards $a_0$ as the probes continue to drift to interstellar space.

Likewise, in the weak-*vsl* limit, we can follow the same logic and derive the first term correction to the doublet forces. In this case equations (29) and (30) become:

$$\nabla \cdot \begin{pmatrix} \dfrac{|\nabla \phi_\Lambda|^2}{a_0} \\ \dfrac{|\nabla \phi_{mond}|^2}{a_o} \end{pmatrix} \cong \nabla^2 \phi_N \begin{pmatrix} \dfrac{3}{2}\dfrac{\nabla \phi_n}{a_0} - \dfrac{3}{4}\left[\dfrac{\nabla \phi}{a_0}\right]^3 \\ 2 + \left[\dfrac{\nabla \phi_n}{a_0}\right]^2 \end{pmatrix} \tag{38}$$

and

$$\begin{pmatrix} \dfrac{|\nabla\phi_\Lambda|^2}{a_0} \\ \dfrac{|\nabla\phi_{mond}|^2}{a_0} \end{pmatrix} \cong \begin{pmatrix} \dfrac{3}{4}\dfrac{|\nabla\phi_n|^2}{a_0}\left[1-\dfrac{1}{4}\left|\dfrac{\nabla\phi_n}{a_0}\right|^2 + \ldots\right] \\ 2\nabla\phi_n\left[1+\dfrac{1}{6}\left|\dfrac{\nabla\phi_n}{a_0}\right|^2 + \ldots\right] \end{pmatrix} \qquad (39)$$

Note, that there is a difference by a factor of two between the dominant eigenforce in the weak-*vsl* limit and the MOND scheme (see black line in figure 2). This is not a big concern, as all the light to mass ratio scalars that were deduced by MOND for the many galaxies are off by a factor of two according to the *vsl* approach. Indeed, this factor of two can come into play (combined with the action of $F_\Lambda$) to alleviate the virial mass discrepancy problem that has been haunting MOND in mass estimates for large galaxy clusters [8, 23].

Equations (29) and (30) condense all non-relativistic gravitational phenomenon from the solar system scale to the universe as a whole in one unified formalism. This is the first theory where dark matter, dark energy, the pioneer anomaly, and Newtonian gravitation are all derived as solutions in the appropriate limits. Moreover, this framework allows for additional predictions as can be seen from equations for both the weak (38 and 39) and strong (36 and 37) *vsl* limits. What remains is a need to find an intuitive explanation for the *vsl*-inertia that emerges from the above description. There are two points to keep in mind: first, Newton's law is modified by allowing a linear mass response to both the eigenforces, thus extending the inertia concept to a complex realm where inertia still carries the same physical meaning, only now it has what could be a hidden dimension. Second, in the strong *vsl*-limit we see a deviation from the canonical second law form, and instead of having a unitary symmetric response, the empirical evidence demands that we break this symmetry. In this case, the physical constraint seems to be that the dominant eigenforce will be so for all ranges of field gradient, and as a result requires that the mass response to this field in this limit will in a sense be renormalized by the field gradient. The mass response for the minor eigenforce does not alter, and as a result it gets renormalized in the opposite sense. Thus, we get a *1/r²* and a constant minor force for this limit.

The mass term as presented in this section relied on our presumed knowledge of the constant or logarithmic pioneer anomaly force. It will take a deeper theoretical reasoning in order to derive this term from basic principles. For the time being, the *vsl*-approach essentially describes the "left-hand" side of the equation, and the "right-hand" side is deduced from observations. In lieu of this, it is important to emphasize that dark energy is a pure *vsl* prediction that does not necessitate the mass response.

# VI. DISCUSSION

The history of development of physics thought can be looked at from a prism of differing points of view. In general, one paradigm is chosen over others, when new observations that confirm predictions, which differentiate it from competing points of view. The fact that quantum mechanics emerges naturally from the *vsl* assumption, does not mean that the varying speed of light/many quantum realities approach is indeed relevant to our universe. It is just another point of view. It, therefore, becomes imperative to examine implications of this hypothesis on known physical observations and pose the following question: can *vsl*-mechanics be used to explain phenomenon that are either currently not understood or explained via as of yet unobserved or unproven mechanism? The simplest and most obvious approach is to apply the formalism initially to classical systems on the very large macroscopic scales. Unlike quantum mechanics, which reduces to the classical equations of motion by averaging out its wave-like effects on scales varying from several atoms and above, here we expect to observe specific effects of this ultra-microscopic assumption (i.e. many quantum realities locally) on the very largest of scales: solar system, galaxy, and universe as a whole. Therefore, *vsl*-mechanics gives us the capability to design stringent tests with very specific and sometimes parameter-less predictions.

When one wants to study motions on large scales, the least phase approximation must be taken for the path integral. In quantum mechanics, this leads immediately to the familiar non-relativistic Newton's laws. In a universe having many quantum realties, we find that Newton's laws are altered. Simply put, *F* does not just equal *ma* – the range and possibility of motion is much wider with immediate implications.

As scientists we take for granted Newton's second law, and in general the concept of inertia. Even though it is one of our most sacred pillars, the notion of inertia and how its tied to the basic fabric of the universe has not received much attention in history. This is surprising, as the very concept of inertia is not a trivial matter. Attempts to alter Newton's second law or the law of inertia, have been rare (two obvious and successful examples are special and general relativity), and many times can be associated with a certain interpretation to Mach's principle: "*The inertia of any system is the result of the interaction of that system and the rest of the universe. In other words, every particle in the universe ultimately has an effect on every other particle*". *Vsl*-mechanics is indeed such a theory, where in this case each particle's motion is effected by the interaction with the sea of universons that perpetuate the motion, thus leading to a more generalized concept of inertia that includes that of Newton but also allows for a continuum of other possibilities.

Another hypothesis that breaks away from Newton's concept of inertia is MOND. About 23 years ago Milgrom introduced this idea as an alternative explanation to the "missing-mass" problem of galaxies and galaxy clusters. The simple MOND formula [5], it turned out was remarkably successful in describing the observed galactic rotation curves, while relying on one fitting parameter (constant mass to light ratio), and a new predicted universal constant termed $a_0$. However, despite its inarguable success, the MOND approach remained on the fringes of conventional physics research programs, and aside for Milgrom and several colleagues the community did not explore this approach seriously. The main reason for this is the fact the MOND, as admitted by Milgrom himself, is not a theory. It is a phenomenological approach that was designed to describe motion of stars and gas on disk galaxies. It is not based on some deep or basic principle, nor is it derived from a larger theoretical construct. Despite their best efforts over the recent two decades Milgrom, Beckenstein and their colleagues have not been able to develop a truly satisfying theory that will derive MOND from basic principles, and thus move it from the fringes to the mainstream of physics research. Now nearly 25 years later, with still not a shred of real evidence for dark matter except for gravitational observations (rotation curves, gravitational lensing, etc.), and the fact that the MOND approach has still not been falsified it may be time to reexamine this issue.

The classical approximation to the *vsl*-mechanics gravitational field theory presented in this paper is the first theory that derives MOND from basic principles. MOND is one particular aspect of the *vsl*-mechanics equations of motion, but it does not capture the particular complexity that is the over-arching feature of the *vsl*-approach. The *vsl*-mechanics deviation from Newtonian mechanics is carried out through the introduction of a complex (non-real) action to describe the motion. The meaning of a complex action at the classical level is highly non-intuitive, and the interpretation made in the context of this work leads to a novel view of inertia. That is, the second law of Newton now reflects an inertia response in the complex plane, thus generating two types of inertia responses: A real (observable) and a complex one. Since any complex observable can be observed in two ways (itself or its complex conjugate), that immediately translates to two observable force responses that are coupled and always appear together. Consequently, any fundamental force field manifests its action on matter (at the classical level) through two forces, a dominant component and a corrective one. The important thing is that the behavior of both forces deviates from the conventional non-complex (real) action prediction. The case treated in this paper deals with the simple Newtonian gravitational field. A non-relativistic, static field. The equations of motion that result from the complex action in this case yield a dominant force which is identical to the MOND force (at the solar system and galaxy level), and a corrective force that together with the dominant force are responsible for two additional gravitational anomalies" a pioneer-like anomaly at the solar system level, and more importantly a repulsive dark-energy like force. Thus, the *vsl*-mechanics approach is the first one to unify all three major gravitational anomalies observed in the past 50 years, namely the missing mass problem, dark energy, and the pioneer anomaly – into one scheme.  It is hard to over-estimate the potential importance of this success to the over-all *vsl* cause.

The successes of *vsl*-mechanics in explaining the three anomalies do not stop here. *Vsl*-mechanics predicts the evolution of the anomaly as the pioneer spaceships continue to escape the solar system – an important prediction which we may be able to test soon. Moreover, the minor force also provides a correction for MOND at the level of galaxies, a test which can help refine the fits for rotation curves, and gain a better estimate for $a_0$. Most importantly, *vsl*-mechanics is a general quantum theory, it is not localized to macroscopic gravity. Therefore, quantum mechanical prediction for the other fundamental forces should emerge naturally. That is, all 4 fundamental force fields should be affected, as *vsl*-mechanics predicts 8 fundamental forces – 2 for each force field. This may not necessarily translate to new elementary particles, but it points to the existence of a higher symmetry principle which is reflected by this concept. A symmetry principle that physics had never before contemplated. From a philosophical perspective, there is much to explore with this concept. For example, how does the *vsl*/universon principle affect the concept of time ? How does special relativity and Lorentz invariance change? What does this mean to the equivalence principle ? There is a deep symmetry principle here that needs to be elucidated.

On the macroscopic scale, the *vsl*-approach will require a new cosmological model – does the universe expand forever? Is the typical representation with mass density still relevant. On the flip side, the approach predicts special physics not just for low accelerations, but also for high ones that reflect the DSR or minimum of absolute distance concept or the discreteness of the universon ideal gas. All of these should be major components of an underlying universon field theory that essentially allows for all the observable phenomenological physics described by QM, classical mechanics, and *vsl*-mechanics to happen.

After examining three large-scale problems it would seem likely that the *vsl*-mechanics approach has relevancy to physical phenomenon observed in our universe. First, quantum mechanics itself is a *vsl*-mechanics prediction. Second, the analysis presented here makes it likely that motion on large solar system, galactic, and cosmological scales can be accounted for adequately, and without resorting to exotic forms of energy or matter in order to explain observations. Finally, the *vsl*-mechanics formalism is a highly predictive one, as it can be easily applied to microscopic problems as well. Thus, we may be able to test predictions in a lab that will corroborate or falsify the galactic scale results. Moreover, because the underlying assumption is the speed of light is not constant on a local scale, this theory may open the door for technologies that will enable the propulsions of material objects not in super-luminal velocities which will conflict with special relativity, but rather in speeds that are larger than $3 \times 10^8$ m/sec, provided that the average speed of light in the local volume of the object accelerated is increased. This last statement may be somewhat in the regime of science fiction, yet *vsl*-mechanics is not. It is a predictive theoretical approach that can be tested in the lab, and until we have no evidence to the contrary, we must consider the possibility that the speed of light is local and that there are many quantum realties available on the ultra-microscopic scales, an assumption which has profound ramifications to ultra-macroscopic phenomenon.

**Figure Captions**

FIG. 1. *Vsl* force functions from weak to strong *vsl* regimes. The figure shows the *vsl* forces as a function of normalized distance. The MOND force function (blue dashed line) as derived for the 1-D case for the weak and strong *vsl* regimes (eqns (106) and (108)) showing the $1/r^2$ and $1/r$ behavior in the strong and weak *vsl* regimes respectively. The red line corresponds to the corrective $F_\Lambda$ force showing the logarithmic and $1/r^2$ behavior in the strong and weak *vsl* regimes respectively. The straight line corresponds to Milgrom's MOND function (eqn (74)), while the green dashed line corresponds to the polar representation of $F_\Lambda$ (eqn (83)) showing that a reasonablly continuous behavior is expected in the transition region from the strong to the weak *vsl* regime. Finally, the black square corresponds to the current measured value for the pionner anomaly, in reasonable agreement with the theoretical prediction.

**FIG. 1.**

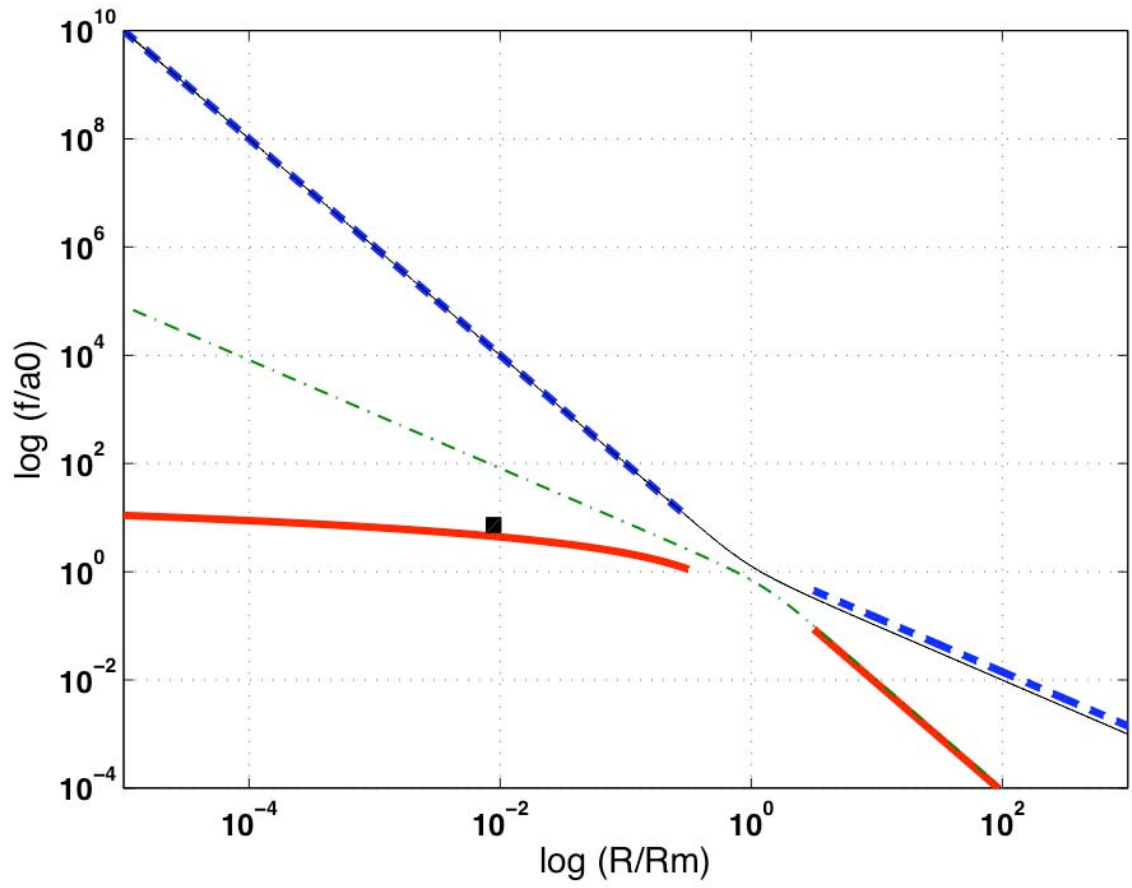

25